\documentclass[twocolumn,times,tighten]{aastex63}

\usepackage{url}
\usepackage{amsmath}
\usepackage{paralist}

\newcommand{\eg}{{e.g.},}
\newcommand{\ie}{{i.e.},}
\newcommand{\cf}{{cf.}}
\makeatletter

\newcommand{\Rmnum}[1]{\expandafter\@slowromancap\romannumeral #1@}
\makeatother

\def\corr#1{{\color{black}{#1}}}
\def\corrtwo#1{{\color{black}{#1}}}

\graphicspath{{./}{figures/}}

\received{\today}
\revised{\today}
\accepted{\today}
\submitjournal{ApJ}

\shorttitle{Localised particle acceleration in the solar corona}
\shortauthors{Long et al.}

\begin{document}

\title{Localised acceleration of energetic particles by a weak shock in the solar corona}

\correspondingauthor{David M. Long}
\email{david.long@ucl.ac.uk}

\author[0000-0003-3137-0277]{David~M.~Long}
\affil{Mullard Space Science Laboratory, UCL, Holmbury St Mary, Dorking, Surrey, RH5 6NT, UK}
\author[0000-0002-6287-3494]{Hamish~A.~S.~Reid}
\affil{Mullard Space Science Laboratory, UCL, Holmbury St Mary, Dorking, Surrey, RH5 6NT, UK}
\author[0000-0001-7809-0067]{Gherardo~Valori}
\affil{Max Planck Institute for Solar System Research, Justus-von-Liebig-Weg 3, 37077 G\"ottingen, Germany}
\author[0000-0002-8806-5591]{Jennifer O'Kane}
\affil{Mullard Space Science Laboratory, UCL, Holmbury St Mary, Dorking, Surrey, RH5 6NT, UK}

\begin{abstract}

Globally-propagating shocks in the solar corona have long been studied to quantify their involvement in the acceleration of energetic particles. However, this work has tended to focus on large events associated with strong solar flares and fast coronal mass ejections (CMEs), where the waves are sufficiently fast to easily accelerate particles to high energies. Here we present observations of particle acceleration associated with a global wave event which occurred on 1 October 2011. Using differential emission measure analysis, the global shock wave was found to be incredibly weak, with an Alfv\'{e}n Mach number of $\sim$1.008--1.013. Despite this, spatially-resolved type~\Rmnum{3} radio emission was observed by the Nan\c{c}ay RadioHeliograph at distinct locations near the shock front, suggesting localised acceleration of energetic electrons. Further investigation using a magnetic field extrapolation identified a fan structure beneath a magnetic null located above the \corr{source} active region, with the erupting CME contained within this topological feature. We propose that a reconfiguration of the coronal magnetic field driven by the erupting CME enabled the weak shock to accelerate particles along field lines initially contained within the fan and subsequently opened into the heliosphere, producing the observed type~\Rmnum{3} emission. These results suggest that even weak global shocks in the solar corona can accelerate energetic particles via reconfiguration of the surrounding magnetic field. 

\end{abstract}


\section{Introduction} \label{sec:intro}

Solar eruptions are the most energetic phenomena occurring in our solar system, emitting bursts of multi-spectral radiation (solar flares), ejecting massive bubbles of plasma into the heliosphere as coronal mass ejections (CMEs) and driving global shock waves that propagate through and restructure the surrounding corona \citep[\cf][]{Warmuth:2015,Long:2017b}. Although initially poorly understood, the advent of synoptic high cadence, multi-wavelength Extreme Ultra Violet (EUV) observations of the low solar corona from the \emph{Solar Dynamics Observatory} \citep[SDO;][]{Pesnell:2012} spacecraft has provided a new insight into global waves in the solar corona. These global waves (formerly known as ``EIT waves'') are identified as bright annulli propagating away from the source of the eruption at typical velocities of 600--730~km~s$^{-1}$ \citep{Nitta:2013,Long:2017b}. While initially characterised using either wave or pseudo-wave interpretations \citep[\cf][]{Warmuth:2015}, \corr{recent work \citep[\cf][]{Long:2017a,Downs:2020} has made it increasingly evident that these features are best interpreted as large-amplitude waves/shocks}. The multi-wavelength observations provided by SDO have also enabled plasma diagnostics of these \corr{perturbations}, showing an adiabatic increase in both temperature and density associated with their passage through the solar corona \citep{Vann:2015,Long:2019,Frassati:2020}. However, in each case these results have underlined the very weak nature of these shocks, finding Mach numbers only slightly greater than 1 \citep[\eg][]{Long:2015,Frassati:2020}.

The very low measured Mach numbers of these global shocks and the \corr{density of the corona} through which they propagate suggests that these global shock waves should be subcritical and hence highly inefficient at accelerating particles. Despite this, previous work has related energetic particles detected in-situ with global waves passing through the predicted footpoint of the connected magnetic field \citep[\cf][]{Rouillard:2012,Prise:2014}. However, these connected observations are relatively rare due to mixing of energetic particles between their origin at the Sun and detection in-situ near the Earth \citep[\cf][]{Laitinen:2019}. 

\begin{figure*}[!ht]
\centering
\includegraphics[width=0.9\textwidth,trim=0 0 0 0,clip=]{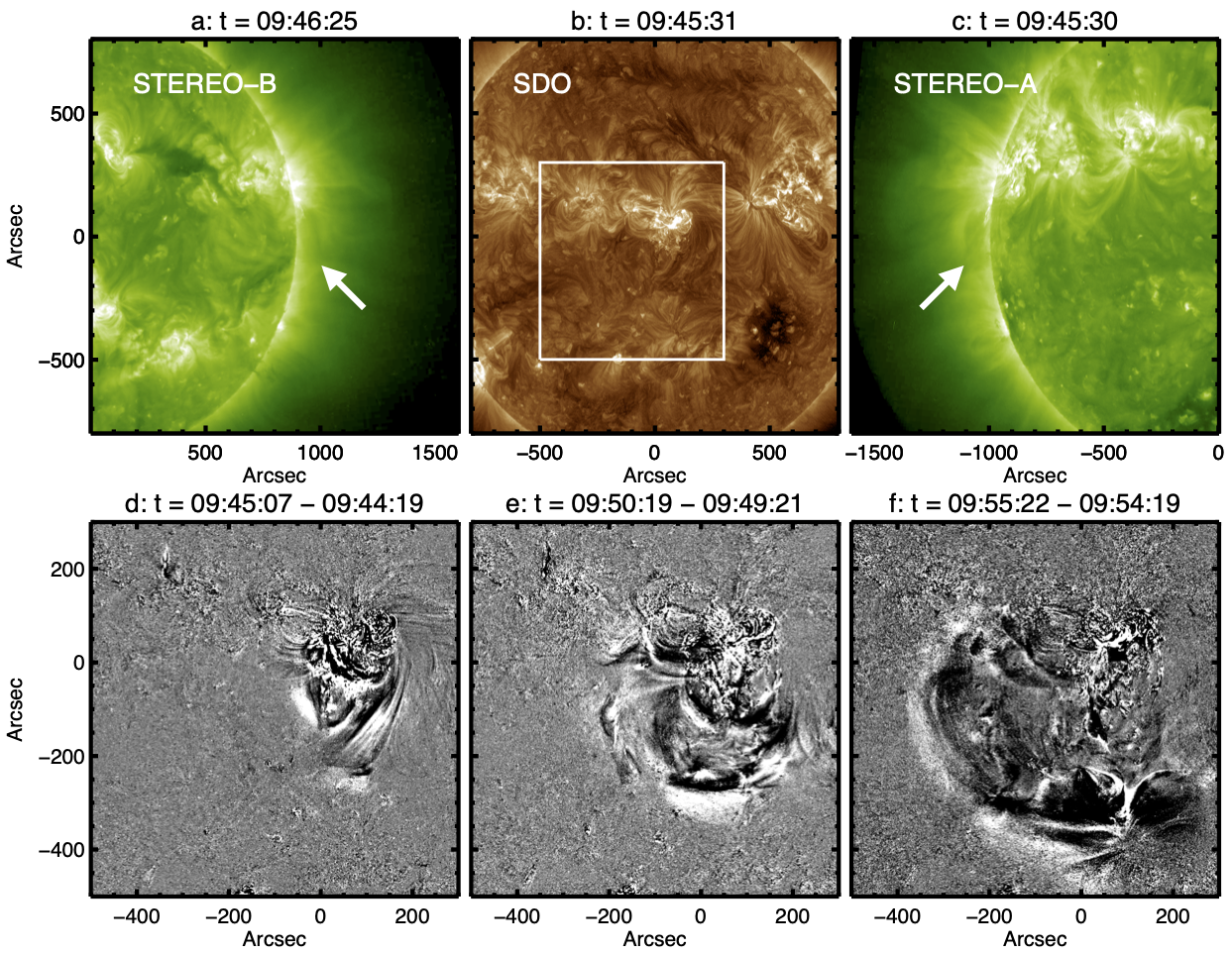}
\caption{\emph{Panel a (c)}: 195~\AA\ images showing the erupting CME bubble (projected leading edge indicated by the white arrow) observed by STEREO-Behind (Ahead). \emph{Panel b}: SDO 193~\AA\ image showing the \corr{source} active region. White box indicates the field of view used for the images in panels d-f and Figure~\ref{fig:radio_context}a-c. \emph{Panels d-f}: SDO 193~\AA\ running difference images (produced by subtracting a following image from a leading image) showing the evolution of the global shock front.
\label{fig:context}}
\end{figure*}

In addition to the direct detection of energetic particles accelerated by global waves, the advent of synoptic high cadence, high resolution observations of the Sun across the electromagnetic spectrum has also provided a unique opportunity to both identify and probe global waves and their relationship to other solar phenomena such as energetic particles. Both \citet{Carley:2013} and \citet{Morosan:2019} identified type~\Rmnum{3} herringbone radio emission as corresponding to rippling in the laterally-propagating front of a CME-driven shockwave (observed in the low corona as a global EUV wave), theorising that this should enable significant particle acceleration by these weak global shocks. However, these signatures were associated with large solar eruptions, typically characterised by X--class flares, very fast CMEs ($v_{CME}\sim$2000~km~s$^{-1}$), and fast, strong global shock wave events. In contrast, most observed global shock wave events are much weaker, suggesting that these very high energy events may not provide the best insight into the processes by which energetic particles are typically accelerated by low energy global shocks in the low corona. \corr{More recent work by \citet{Morosan:2020} has shown evidence of extended radio emission produced by a solar eruption with an associated global wavefront weaker than the events studied by \citet{Carley:2013} and \citet{Morosan:2019}. However, this event erupted from the backside of the Sun and could only be studied from Earth once it appeared above the solar limb, thus complicating a full analysis of the global wavefront.}

Here we use a combination of multi-spectral observations and modelling to investigate the evolution of a weak global shock wave in the solar corona associated with spatially resolved type~\Rmnum{3} radio emission. The event and associated data sets are described in Section~\ref{s:obs}. The results are presented in Section~\ref{s:results}, examining the evolution of the radio emission (Sect.~\ref{ss:radio}), plasma signatures (Sect.~\ref{ss:plasma}), and the magnetic field (Sect.~\ref{ss:bfield}), before these diagnostics are discussed and some conclusions drawn in Section~\ref{s:disc}.

\section{Observations and Data Analysis}\label{s:obs}

On 2011~October~1, a M1.2 solar flare erupted from NOAA Active Region (AR) 11305, near the centre of the solar disk as seen from Earth. It was associated with an Earth-directed CME, and was well observed at EUV wavelengths by the Atmospheric Imaging Assembly \citep[AIA;][]{Lemen:2012} onboard the SDO spacecraft near Earth and the Extreme UltraViolet Imagers \citep[EUVI;][]{Wuelser:2004} onboard the two Solar Terrestrial Relations Observatory \citep[STEREO;][]{Kaiser:2008} Ahead and Behind spacecraft, which were 104.3 and 97.5 degrees ahead of and behind the Earth respectively on its orbit around the Sun. The eruption was also associated with type~\Rmnum{2} and \Rmnum{3} radio emission observed by the Nan\c{c}ay RadioHeliograph \citep[NRH;][]{Kerdraon:1997} and the Nan\c{c}ay Decameter Array \citep[NDA;][]{Lecacheux:2000}. As it erupted, the CME drove a global shock wave which propagated primarily towards the southeast of the solar disk from the \corr{source} active region. 

\begin{figure*}[!ht]
\centering
\includegraphics[width=0.92\textwidth,trim=0 0 0 0,clip=]{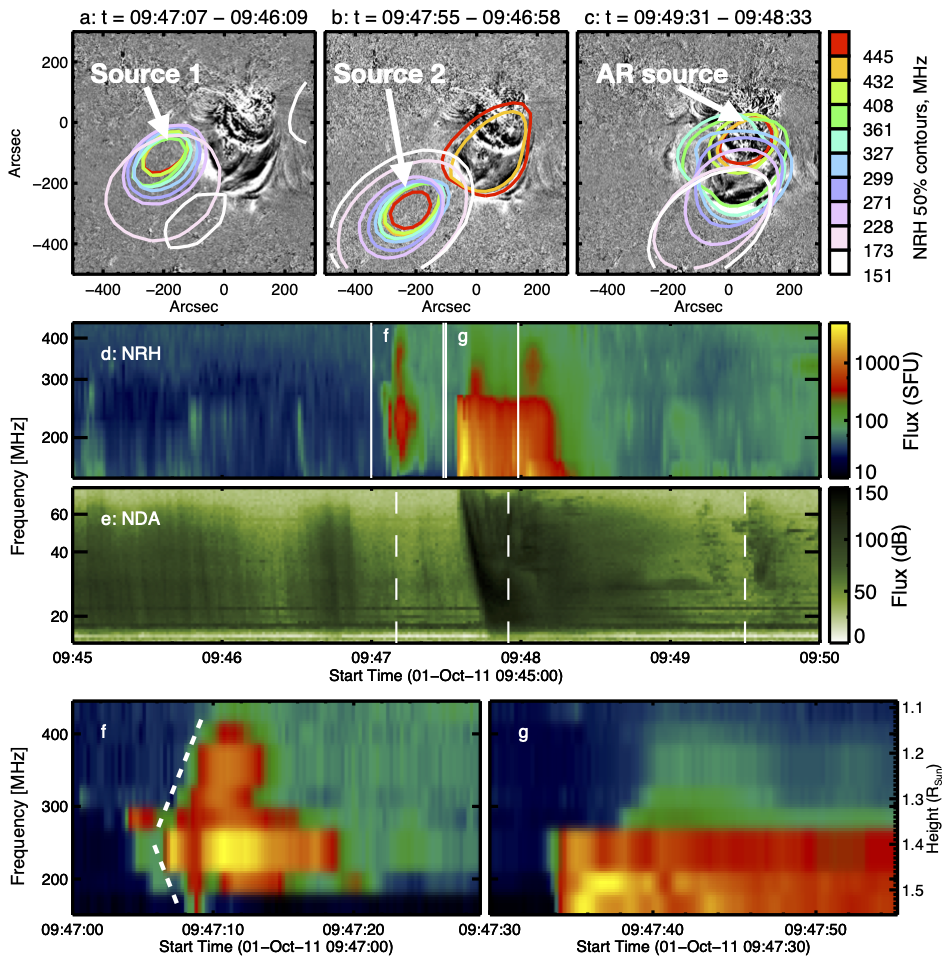}
\caption{\emph{Panel a-c}: SDO 193~\AA\ running difference images showing the evolution of the global shock front. Contours show the radio data from the Nan\c{c}ay radioheliograph (NRH), with colour indicating frequency. In each case the \corr{times of the images used to produce the difference image} is shown in the title. \corr{\emph{Panel~d}: NRH Radio spectrum, with vertical lines indicating the regions of the spectrum used to produce panels f \& g. \emph{Panel~e}: NDA radio spectrum, with dashed vertical lines indicating} the times of the radio contours shown in panels a-c. \emph{Panel f (g)}: Zoom-in on the NRH observations of the first (second) radio burst without any interpolation in frequency. \corr{Dashed white lines in panel~f indicate the bifurcation of the radio signal as noted in the text.}
\label{fig:radio_context}}
\end{figure*}

The event was well observed by the STEREO and SDO spacecraft, with the data in each case reduced and processed using the standard SolarSoftWare routines \citep{Freeland:1998}. Figure~\ref{fig:context} shows the erupting CME as observed by the two STEREO spacecraft (panels~a \& c), with panel~b showing the \corr{source} active region as seen by the SDO spacecraft near Earth. The global wave can be identified as the white feature propagating away from the source active region in Figure~\ref{fig:context}d-f, which highlight the feature using difference images (produced by \corr{subtracting an image from another taken at a later time}). The associated radio emission is shown in Figure~\ref{fig:radio_context}. Panels~a-c \corr{in Figure~\ref{fig:radio_context}} show the location and frequency of the radio emission at 50~\% of the peak intensity, while panels~d \& e show the radio spectra observed by NRH and NDA respectively. The three vertical dashed lines \corr{in panel~e} correspond to the times of the images shown in panels~a-c. Panels~f \& g show a zoom-in on the NRH observations of the radio bursts \corr{(as indicated in panel~d)} offset from the active region identified in panels~a \& b without any interpolation in frequency.

\begin{figure*}[!ht]
\centering
\includegraphics[width=0.99\textwidth,trim=0 0 0 0,clip=]{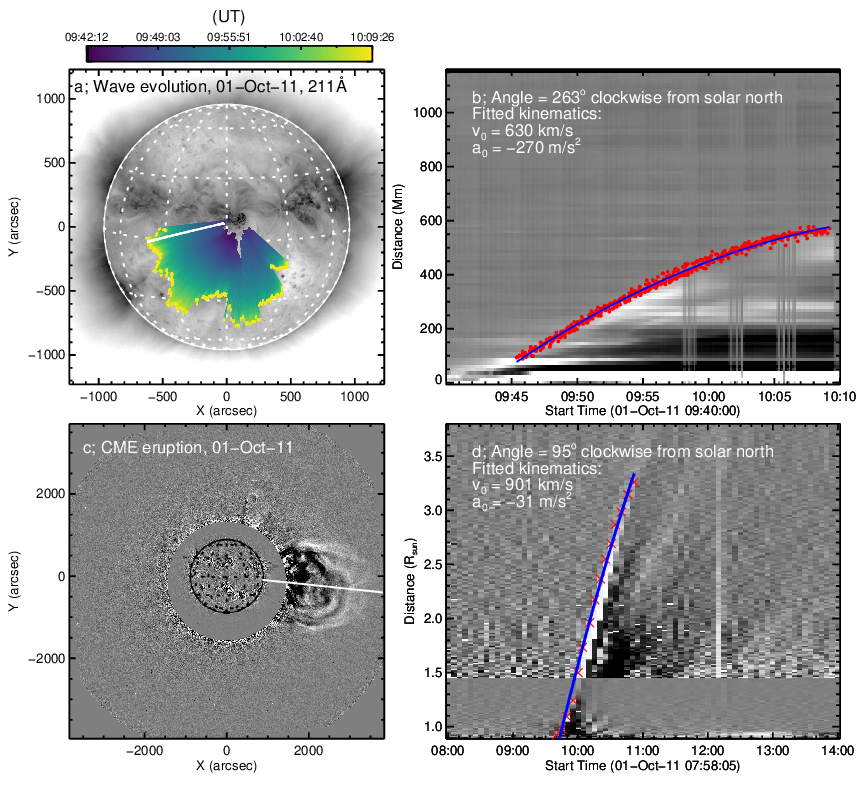}
\caption{\emph{Panel~a}; The evolution of the global wave as estimated by eye using a stack plot approach. \emph{Panel~b}; The stack plot along the angle 263$^{\circ}$ clockwise from solar north (as indicated by the white line in panel a). Red dots indicate the points identifying the leading edge of the wavefront along the arc, with the blue line indicating a quadratic function fitted to the red points. The fitted initial velocity and constant acceleration for this arc sector are given in the legend. \corr{Note that the vertical grey lines are due to intensity changes as a result of varying instrument exposure time, but do not affect the identification of the wavefront.} \emph{Panel~c}; The erupting CME observed using STEREO-B EUVI and Cor-1. \emph{Panel~d}; the intensity stack plot along the white line in panel~c used to estimate the kinematics of the erupting CME. Blue line indicates the quadratic fit to the red data-points with the fitted initial velocity and constant acceleration along this cut given in the legend.
\label{fig:wave_evol}}
\end{figure*}

The evolution of the global EUV wave was tracked on-disk using an intensity stack plot approach \citep[cf.][]{Long:2019} in the 211~\AA\ passband observed by SDO/AIA \citep[as this passband has been shown to be optimal for identifying and tracking global EUV waves, \eg][]{Long:2014}. A series of intensity stack plots similar to that shown in Figure~\ref{fig:wave_evol}b were used to track the temporal evolution of the global EUV wave, with the wave identified by eye 5 independent times each using 100 points along a defined angle to minimise the associated uncertainty. All 500 points (as shown by the red dots in Figure~\ref{fig:wave_evol}b) were then used to determine the kinematics of the wavefront along that angle using a quadratic fit (indicated by the blue line in Figure~\ref{fig:wave_evol}b, with the fit parameters for this angle given in the panel legend). The process was repeated for each angle, enabling the spatial and temporal evolution of the global wave to be determined as shown in Figure~\ref{fig:wave_evol}a, with the white line here corresponding to the intensity stack plot shown in panel~b. 

The CME was observed by both STEREO spacecraft to erupt approximately in the plane of sky from the Earth-directed limb, enabling it to be tracked through the EUVI and Cor-1 fields of view. Figure~\ref{fig:wave_evol}c shows the STEREO-B EUVI and Cor-1 fields-of-view, with the white solid line indicating the angle used to derive the intensity stack plot shown in Figure~\ref{fig:wave_evol}d. This angle (at 95$^{\circ}$ clockwise from solar north) was chosen as it cut through the centre of the plane-of-sky CME structure as it propagated through the field of view of both EUVI and Cor-1. The intensity along this line was then taken for each time step, with the bright edge of the CME identified manually and fitted using a quadratic function to derive its kinematics \citep[following the stack plot approach described by][]{okane:2019}. 

As shown in Figure~\ref{fig:wave_evol}a, the global wave exhibited clearly anisotropic evolution, consistent with the \corr{initially} southward directed eruption of the CME observed by both STEREO spacecraft and shown in Figure~\ref{fig:context}, and similar to the events previously studied by \citet{Long:2019}. \corr{However, it is worth noting that by the time the CME was observed by Cor-1 onboard both STEREO spacecraft, it had become much more symmetric (as shown in Figure~\ref{fig:wave_evol}c), suggesting that by this time it was no longer constrained by the coronal magnetic field.} The global wave was observed to propagate with an average initial velocity of 733~km~s$^{-1}$ and acceleration of $-427$~m~s$^{-2}$, making it slower than previously studied global shock wave events \citep{Long:2015,Long:2017a}, but still faster than global shocks observed using the previous generation of solar instrumentation \citep[\eg\ SOHO/EIT \& STEREO/EUVI; \cf][]{Thompson:2009,Muhr:2014}. The CME was found to have a velocity of 901~km~s$^{-1}$ with an associated acceleration of $-31$~m~s$^{-2}$ in the direction of Earth.

The NRH data at 0.25~s cadence was processed using the standard techniques and used for performing imaging spectroscopy between 450~MHz to 150~MHz to examine the type~\Rmnum{3} radio emission associated with this global wave event \citep[see][for a recent review]{Reid:2020}. NDA spectroscopy data at 1~s cadence was used to examine the type~\Rmnum{3} radio emission at frequencies between 80-14~MHz. A narrow type~\Rmnum{3} radio burst can be observed in the radio spectrum shown in Figure~\ref{fig:radio_context}d \& \ref{fig:radio_context}f at t$\sim$09:47:10~UT, corresponding to source~1 identified in Figure~\ref{fig:radio_context}a. However, this type~\Rmnum{3} burst is only observed by NRH, indicating that it does not move to lower frequencies and hence higher altitudes. In contrast, an extended type~\Rmnum{3} burst can be identified starting at t$\sim$09:47:39~UT (Figure~\ref{fig:radio_context}g), which is observed by both NRH and NDA. Figure~\ref{fig:radio_context}b indicates that some of this emission comes from source~2, with a significant amount of emission also coming from the source active region (shown in Figure~\ref{fig:radio_context}c). Although not seen in the time frame shown here, a type~\Rmnum{2} radio burst was also observed associated with this event \corr{starting at 09:55~UT}, indicating the presence of a CME-driven shock in the outer corona.

\section{Results}\label{s:results}

The radio emission associated with this event indicates the presence of a CME-driven shock (type~\Rmnum{2} emission) and accelerated electrons (type~\Rmnum{3} emission), both of which have previously been associated with global EUV waves \citep[\eg][]{Harvey:1974,Warmuth:2004a,Warmuth:2004b,Carley:2013,Morosan:2019}. In this case, the type~\Rmnum{3} emission was identified using imaging spectroscopy as coming from three distinct locations; the \corr{source} active region and two sources ahead of the leading edge of the propagating wave front. These regions are identified by the white arrows in panels~a-c of Figure~\ref{fig:radio_context}; panels~a \& b show the sources associated with the propagating wave front (source 1 \& 2) while panel~c shows the active region source. Although sustained type~\Rmnum{3} emission coming from the \corr{source} active region can be explained by flare-driven electron acceleration, the origin of the short-lived emission coming from regions 1 \& 2 identified in Figure~\ref{fig:radio_context}a \& b is less clear and requires further investigation.

\begin{figure*}[!ht]
\centering
\includegraphics[width=0.99\textwidth,trim=0 0 0 0,clip=]{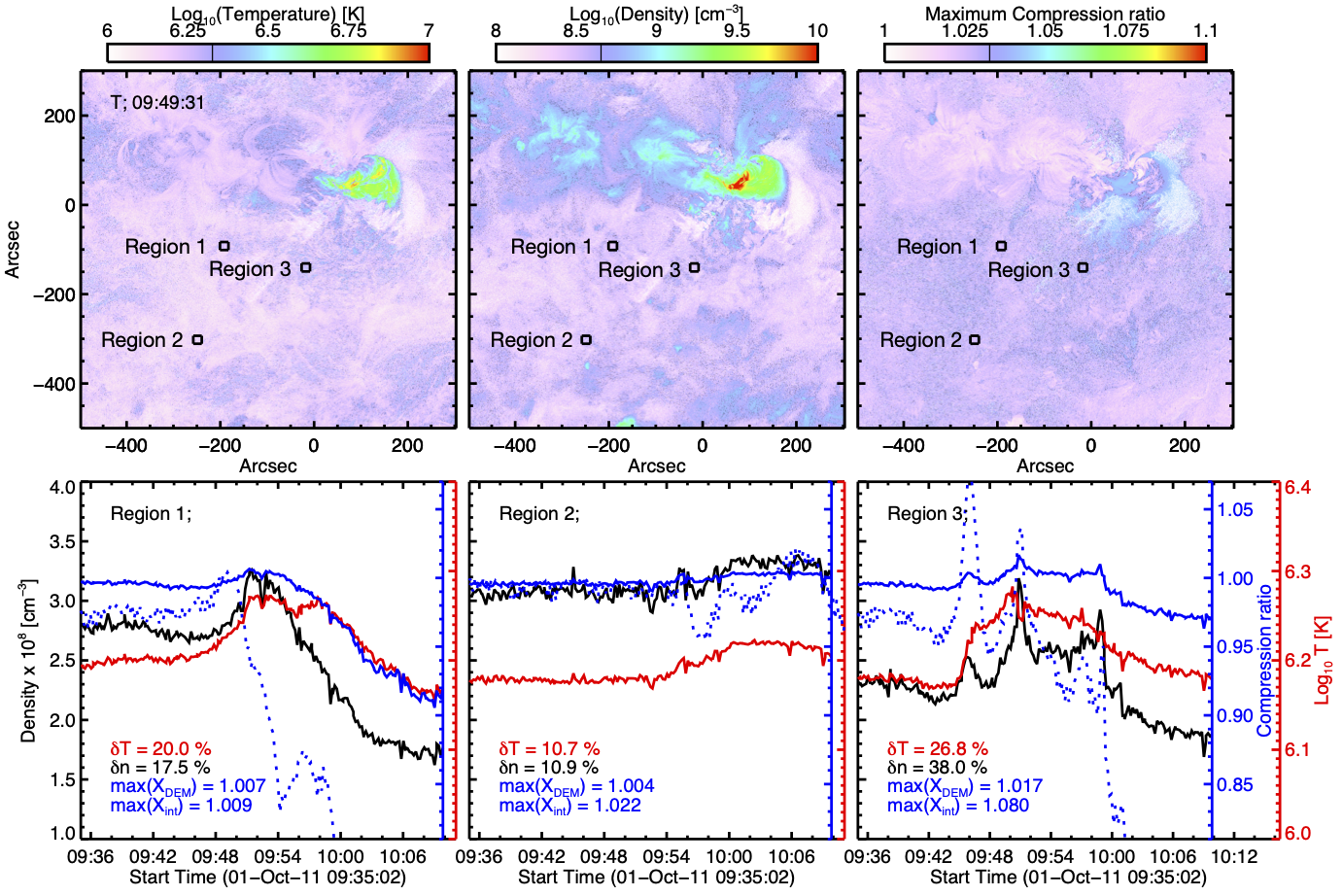}
\caption{\emph{Top row}: Average temperature (\emph{left}), density (\emph{centre}), and maximum compression ratio (\emph{right}) at 09:49:31~UT, each derived using the regularised inversion technique of \citet{Hannah:2013}. Boxes indicate the regions that were used to derive the average densities, temperatures and compression ratios. \emph{Bottom row}: The temporal variation of the DEM-inferred density (black), temperature (red) and compression ratio (blue) in Region~1 (left), Region~2 (centre), and Region~3 (right). In each case, the relative start to peak changes in density, temperature, and maximum compression ratio values derived using the DEM (solid blue line) and 193~\AA\ intensity (dashed blue line) approaches are indicated in the legend.
\label{fig:temp_dens_evol} }
\end{figure*}

\subsection{Radio emission}\label{ss:radio}

\corr{Panels~d \& e of Figure~\ref{fig:radio_context} show the dynamic spectra for the NRH and NDA instruments respectively, in each case derived from the imaging spectroscopy shown in Figure~\ref{fig:radio_context}a-c using a field-of-view from ($-670'',-670''$) in the bottom left to ($287'', 287''$) in the top right. Figure~\ref{fig:radio_context}f, g} show zoomed-in sections of the NRH radio spectrum at the times of the radio bursts identified as occurring away from the \corr{source} active region (in both cases with no interpolation across frequency). It is worth noting that whereas the zoom-in of source~1 (Figure~\ref{fig:radio_context}f) shows a \corr{clear} bifurcation of the radio emission (indicating electrons accelerated both towards and away from the Sun), the zoom-in of source~2 (Figure~\ref{fig:radio_context}g) \corr{shows clear emission drifting to lower frequency but very weak emission drifting to higher frequency, indicating that the electrons were primarily propagating away from the Sun. Note that due to the very weak nature of the downward emission, it was not possible to estimate a drift rate for it.}

We can identify properties of the particle acceleration region within \corr{source~1} (Figure~\ref{fig:radio_context}f) using the type~\Rmnum{3} radio emission. The density within the acceleration region can be estimated using the frequency at which the bi-directional type~\Rmnum{3} bursts start. Assuming harmonic emission \corr{\citep[due to the diffuse nature of the radio emission making it difficult for fundamental emission to escape the corona at higher frequencies, \eg][]{Reid:2020}}, this provides an electron density of $\sim3.3\times10^8$~cm$^{-3}$. The height of the acceleration region can then be estimated by assuming an electron density model and we use a 4x Saito model \citep{Saito:1977}. As our radio source was on-disk, we could not use the source centroids to estimate the density model and so we chose the 4x multiplier to match the frequency of the bi-directional electron beams with the expected height range of the global EUV shock \citep[$\sim$70--100~Mm above the photosphere, \cf][]{Patsourakos:2009,Kienreich:2009}. \corr{While this assumption introduces a source of error, active regions typically have higher coronal densities than the quiet Sun which the standard Saito model represents.} Assuming the 4x Saito model, the altitude at which the bifurcation occurs and hence where the particles are accelerated, corresponds here to $\sim$87~Mm above the photosphere for source~1. The relatively small frequency range where the type~\Rmnum{3} emission starts indicates that the acceleration region is likely small in size, on the order of megametres or less, as the beam must quickly become unstable to Langmuir waves which produce the radio emission \citep{Reid:2014}.

The velocity of the accelerated electron beams can be estimated using the assumed density model. A linear fit was made to the times of peak flux as a function of altitude, corresponding to the different frequencies observed by the NRH. The 0.25~second time resolution of the NRH necessitated that the times of peak flux be estimated by fitting the light curves using a non-symmetric Gaussian function. The resultant velocities of the associated electron beams were 69~Mm~s$^{-1}$ and 82~Mm~s$^{-1}$
for the beams moving away from and towards the Sun, respectively. This equates to energies of $\sim20$~keV and $\sim28$~keV respectively. However, the electron beams consist of electrons with a distribution of velocities, travelling in a beam-plasma structure due to wave-particle interactions \citep[\eg][]{Reid:2018}. If we assume that the maximum electron velocity within the beam is 1.5x the derived velocity from the type~\Rmnum{3} emission, the maximum energies of the electrons within the beams moving away from and towards the Sun increase to 44~keV and 63~keV respectively. It must also be noted that multiple crossings of the shock front would be required to achieve these energies \citep[\cf][]{Ball:2001}. 

For \corr{source~2} (Figure~\ref{fig:radio_context}g), the downward emission is much weaker, with the upward emission originating at $\sim$230--270~MHz (corresponding to a density of $2\times10^8$~cm$^{-3}$ assuming harmonic emission). Although it was not possible to estimate a beam velocity for the upward propagating beam due to a lack of data-points in the NRH observations, clear type~\Rmnum{3} emission corresponding to source~2 is also observed in the NDA dynamic spectrum (Figure~\ref{fig:radio_context}e). The 1~s time resolution of the NDA necessitated fitting the peak intensity as a function of frequency for each time bin using an asymmetric Gaussian so that the beam velocity could be found. A velocity of 79~Mm~s$^{-1}$ was found, similar to that of source~1 and with comparable energy. As for source~1, it was also possible to estimate the height of the particle acceleration for source~2, corresponding to $\sim$137-185~Mm above the photosphere for emission originating between 230-270~MHz.

\subsection{Plasma diagnostics}\label{ss:plasma}

Next, the evolution of the coronal plasma associated with the passage of the global EUV shock front was further examined to identify variations in temperature and density. As discussed by \citet{Long:2019}, the spatial, temporal and spectral resolution provided by SDO/AIA has enabled the development of a series of techniques designed to quantify the differential emission measure (DEM) of EUV plasma in the low solar corona \citep[see, \eg\ the work by][for a number of different techniques designed to compute DEMs using observations from SDO/AIA]{Hannah:2013,Plowman:2013,Cheung:2015,Morgan:2019}. The DEM $\phi(T)$ is defined as,
\begin{equation}
    \phi(T) = n_e^2(T)\frac{dh}{dT}, \label{eqn:dem}
\end{equation}
where $n_e$ is the density. Each of these techniques offers a different approach to solving this ill-posed problem, providing an opportunity to derive the DEM. This can then be used to derive the DEM-weighted density and temperature using the approach of \citet{Vann:2015,Long:2019}. As noted by \citet{Cheng:2012}, the DEM-weighted density is defined as,
\begin{equation}
    n_e = \sqrt{\frac{\int\phi(T)dT}{h}}, \label{eqn:dens}
\end{equation}
where \corr{$h$} is the plasma scale height. Similarly, the DEM-weighted temperature is defined as,
\begin{equation}
    T = \sqrt{\frac{\int\phi(T)TdT}{\int\phi(T)dt}}. \label{eqn:temp}
\end{equation}

To quantify the role of the global wave in the origin of the type~\Rmnum{3} radio emission labelled as sources 1 \& 2 in Figure~\ref{fig:radio_context}, the DEM of the plasma observed by SDO/AIA was estimated using the Regularised Inversion technique of \citet{Hannah:2013}. This technique combines the 6 coronal EUV passbands observed by SDO/AIA (94, 131, 171, 193, 211, 335~\AA), enabling an estimate to be made of the emission measure of the region indicated by the white box in Figure~\ref{fig:context}b. The DEM-averaged temperature of this region was estimated using Eqn.~\ref{eqn:temp}, and is shown in Figure~\ref{fig:temp_dens_evol}a at 09:49:31~UT. The DEM-weighted density was estimated using Eqn.~\ref{eqn:dens} with a scale height of 90~Mm \citep[\cf][]{Patsourakos:2009,Vann:2015,Long:2019}, and is shown in Figure~\ref{fig:temp_dens_evol}b at 09:49:31~UT. In both panels~a \& b of Figure~\ref{fig:temp_dens_evol}, the temperature and density behave as expected, showing increased values in the core of the \corr{source} active region, with lower values in the quiet Sun where the wave was observed to propagate. 

Three regions of interest were identified to examine the temporal variation of the temperature and density; regions~1 (corresponding to radio source 1), 2 (corresponding to radio source 2) and 3 (representing the background quiet Sun). The variation in temperature (density) with time in each region is shown in red (black) in the lower row of Figure~\ref{fig:temp_dens_evol}, which shows the variation in regions 1, 2, and 3 from left to right. Region~1, corresponding to radio source~1, exhibits increases of $\sim$20~\% and 17.5~\% above the pre-event temperature and density values respectively, while region~2 which is located further from the source active region, exhibits a much smaller increase of $\sim$11~\% above the pre-event level in both temperature and density. Region~3, which is closer to the source active region than both regions~1 \& 2 but does not have a corresponding radio source, exhibits higher increases in both temperature ($\sim$27~\%) and density ($\sim$38~\%). However, the density variation in region~3 is bursty and highly variable while the temperature is comparably smooth, with observations \corr{from the 304~\AA\ passband} indicating that this bursty behaviour is due to filamentary material associated with the core of the CME passing through the field of view.

The variation in density and temperature above the pre-event values observed in regions~1, 2, \& 3 is proportional to their distance from the \corr{source} active region, consistent with a radially-expanding shock front \citep[\cf][]{Downs:2020}. However, the variation in temperature observed here is much larger than that observed for other, comparable global shock wave events \citep[\cf][]{Vann:2015,Long:2019}, suggesting a significant role for non-adiabatic processes. In particular, for both region~1 and 3 (bottom-left and -right panels of Figure~\ref{fig:temp_dens_evol}), the relative temperature and density increases are comparable, with sudden density increases at the front of the wave followed by drop below the pre-event value while the temperature remains enhanced for a longer duration than the density before dropping back to its pre-event value. This suggests extended heating associated with the passage of the global wave front in each location close to the origin. In contrast, region~2 exhibits gradual and comparable increases in both density and temperature, with the return to the pre-event value occurring after the time-range examined here.

The DEM can also be used to estimate the density compression ratio and hence strength of the global shock wave. Although this is typically estimated using radio data by examining the difference in frequency between the fundamental and harmonic frequencies of the shock, the type~\Rmnum{2} emission observed here corresponds to the leading edge of the CME driven shock rather than the laterally driven global wave shock, and could therefore provide an inaccurate estimate of the shock strength. Instead, the density compression ratio can be estimated by examining the change in EUV intensity and/or DEM \citep[\cf][]{Muhr:2011,Zhukov:2011,Ma:2011,Kozarev:2011,Frassati:2019,Long:2019,Frassati:2020}. The EUV density compression ratio ($X$) was estimated here using both the intensity ratio approach of \citet{Zhukov:2011} and the DEM ratio approach of \citet{Frassati:2019}.

As described by \citet{Frassati:2019}, the density compression ratio can be estimated by examining the variation between pre- and post-event EM using the equation,
\begin{equation}
    X = \sqrt{\frac{EM_{U} - EM_{D}}{P_{U}}+1},
\end{equation}
where $EM_{U}$ and $EM_{D}$ are the upstream (\ie\ ahead of the shock) and downstream (\ie\ behind the shock) emission measure respectively and $P_{U}$ is the contribution to the pre-event emission measure from coronal plasma compressed by the passage of the global wave (\ie\ the plasma located between the points $L_{1}$ and $L_{2}$ along the line of sight). $P_{U}$ is defined as,
\begin{equation}
    P_{U} = L <n_{e, U}^{2}>_{LOS} \textrm{ (with } L = L_{2}-L_{1}).
\end{equation}

Alternatively, the density compression ratio can be estimated using the approach of \citet{Zhukov:2011}. Here, the density compression ratio can be related to the intensity ratio of the 193~\AA\ passband via the approximation,
\begin{equation}
    \frac{n}{n_{0}} = \sqrt{\frac{I}{I_{0}}},
\end{equation}
where $I_{0}$ and $n_{0}$ are the intensity and density, respectively, prior to the passage of the global wave. This approach only considers plasma in the temperature range of the 193~\AA\ passband ($T\sim1$~MK), and as a result is much more sensitive to the convolution of the timescale for ionisation of the plasma with the timescale for increased plasma emission due to the passage of the shock \citep[as discussed by][]{Ma:2011}. 

The variation in density compression ratio in each of the three regions estimated using both approaches is shown in the bottom row of Figure~\ref{fig:temp_dens_evol}. Here, the solid blue line shows the density compression ratio estimated using the EM technique of \citet{Frassati:2019}, with the top right panel showing the maximum compression ratio across the whole field of view estimated using this technique. The \corr{dotted} blue line in the bottom row of panels of Figure~\ref{fig:temp_dens_evol} shows the density compression ratio estimated using the 193~\AA\ intensity approach of \citet{Zhukov:2011}. It is clear that while both compression ratio techniques provide similar results, there are distinct differences between them, with the compression ratio derived using the 193~\AA\ intensity approach exhibiting much more variability than the EM approach.

The evolution in density, temperature, and compression ratio shown in Figure~\ref{fig:temp_dens_evol} indicate that the global wave was freely propagating and very weakly shocked at the time and location of the observed type~\Rmnum{3} radio emission. \corr{The estimated variation in density provides an excellent match to the radio emission, with the estimated height of the global wave and the starting frequency of the radio emission broadly consistent.} The highly specific origins of the type~\Rmnum{3} emission suggests that the shock wave may have locally created the conditions to accelerate the electrons that produced the observed emission. By combining the wave speed with the estimated density compression ratios in these locations, it is possible to infer the Mach number of the observed global wave. 

Using the approach described in Section~\ref{s:obs}, the speed of the wave was determined at regions~1, 2, \& 3 as identified in Figure~\ref{fig:temp_dens_evol}. The global wave was found to have a velocity of v = $\sim$574~km~s$^{-1}$ at region~1 (corresponding to temperature and density increases of $\sim$20~\% and 17.5~\% respectively), a velocity of v = $\sim$410~km~s$^{-1}$ at region~2 (corresponding to temperature and density increases of $\sim$11~\%) and a velocity of v = $\sim$408~km~s$^{-1}$ at region~3 (corresponding to temperature and density increases of $\sim$27~\% and $\sim$38~\% respectively). These observations are consistent with a shock wave propagating freely through the low solar corona.

\begin{deluxetable}{cccc}[!t]
\tablecaption{Plasma \& wave properties} \label{tbl:plasma}
\tablecolumns{4}
\tablewidth{0pt}
\tablehead{
\colhead{Property} & \colhead{Region 1} & \colhead{Region 2} & \colhead{Region 3}
}
\startdata
$v_{sh}$ (km~s$^{-1}$) & 574.30 & 410.99 & 407.83 \\
$c_{s}$ (km~s$^{-1}$) & 292.02 & 291.65 & 291.66 \\
$M = v_{sh}/c_{s}$ & 1.967 & 1.409 & 1.398 \\
$v_{A}$ (km~s$^{-1}$, $\theta=0^{\circ}$) & 572.42 & 407.61 & 407.24 \\
$M_{A} = v_{sh}/v_{A}$, $\theta=0^{\circ}$) & 1.003 & 1.008 & 1.001 \\
$v_{A}$ (km~s$^{-1}$, $\theta=90^{\circ}$) & 571.24 & 405.75 & 406.77 \\
$M_{A} = v_{sh}/v_{A}$, $\theta=90^{\circ}$) & 1.005 & 1.013 & 1.003 \\
$B$ (G) & 3.5 & 2.5 & 2.5 \\
\enddata
\end{deluxetable}

The local sound speed ($c_s = (\gamma k_B T/m)^{1/2}$) and thus hydrodynamic Mach number ($M = v_{sh}/c_{s}$) of the shock were estimated in each region of interest using the temperature derived above, and are listed in Table~\ref{tbl:plasma}. It is clear that the wave was weakly shocked with a hydrodynamic Mach number of $\sim1.4$--$1.9$ depending on location. 

The Alfv\'{e}n Mach number ($M_{A} = v_{sh}/v_{A}$) and hence Alfv\'{e}n speed $v_A$ can be estimated from the density compression ratio ($X$) using the approach of \citet{Vrsnak:2002} via the relationship,
\begin{multline}
    (M_{A}-X)^2[5\beta X + 2M_{A}^{2}\textrm{cos}^{2}\theta (X-4)] \\
    + M_{A}^{2}X \textrm{sin}^{2}\theta[(5+X)M_{A}^{2} +2X(X-4)]=0,
\end{multline}
where $\beta$ is the ratio of plasma to magnetic pressure, and assuming an adiabatic index $\gamma=5/3$. In the case $\beta\rightarrow0$, this reduces to,
\begin{equation}
    M_{A} = \sqrt{\frac{X(X+5)}{2(4-X)}},
\end{equation}
for a perpendicular shock ($\theta = 90^{\circ}$), and,
\begin{equation}
    M_{A} = \sqrt{X},
\end{equation}
for a parallel shock ($\theta = 0^{\circ}$). The resulting Alfv\'{e}n speeds and Alfv\'{e}n Mach numbers for $\theta=90^{\circ}$ and $\theta=0^{\circ}$ derived using the DEM density compression values given in Figure~\ref{fig:temp_dens_evol} are shown in Table~\ref{tbl:plasma}. While there is very little difference between the derived Alfv\'{e}n speed values for a perpendicular and parallel shock in each of the regions studied here, it is worth noting that perpendicular shocks are much more efficient at accelerating particles than parallel shocks. 

The derived Alfv\'{e}n velocity can also be used to estimate the magnetic field strength, $B$, of the three regions using the equation,
\begin{equation}
    B = v_A \sqrt{4 \pi n_e m_i}.
\end{equation}
The values of the Alfv\'{e}n speed given in Table~\ref{tbl:plasma} give corresponding magnetic field strength values of 3.5~G, 2.5~G, and 2.5~G for regions~1, 2, and 3 respectively. These estimates are consistent with previous estimates of the coronal magnetic field strength made using global EUV waves \citep[\cf][]{West:2011,Long:2013}.

\begin{figure*}[!ht]
\begin{center}
\includegraphics[width=\textwidth, trim=0 0 0 0,clip=]{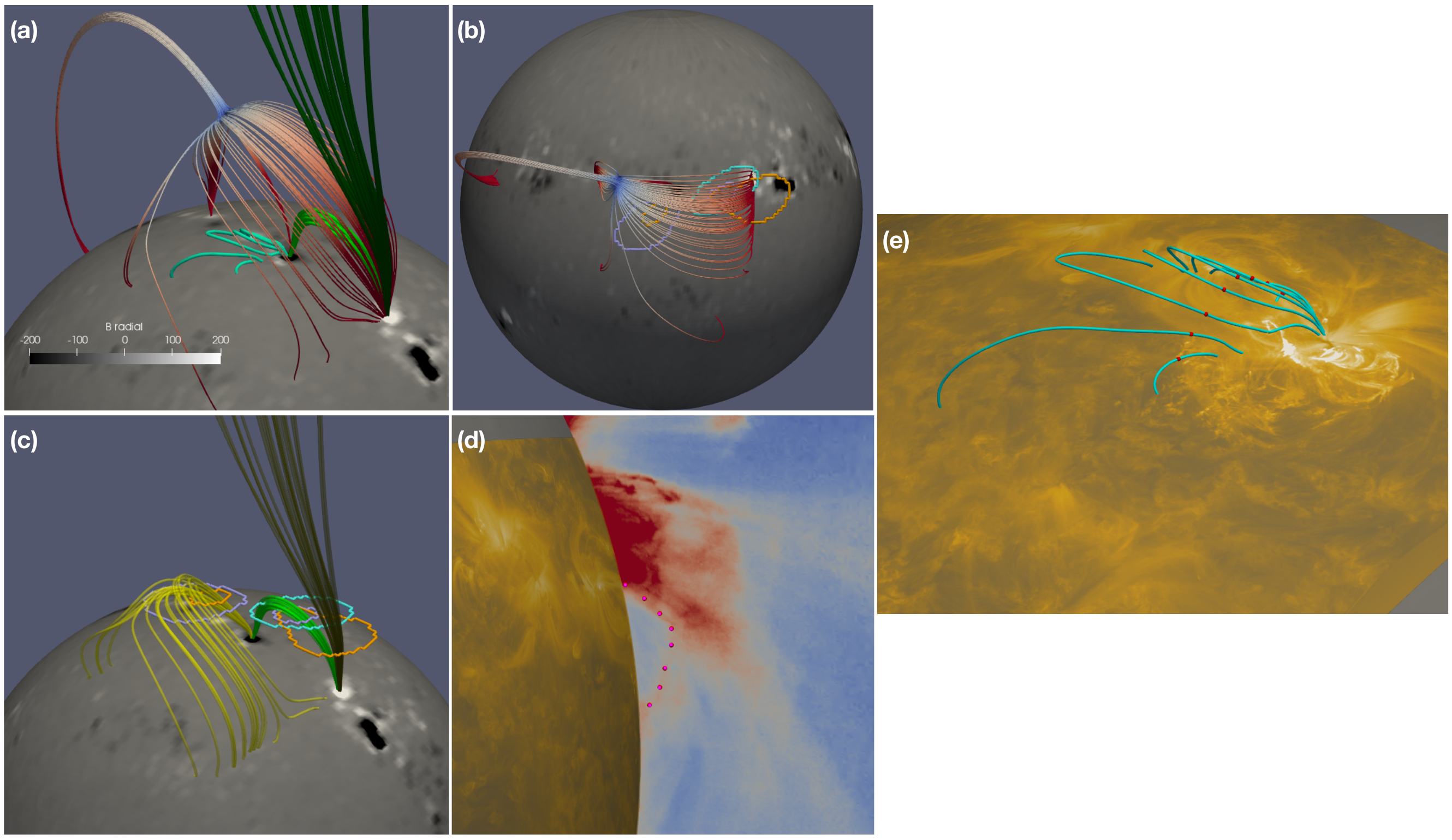}
\caption{\corr{PFSS extrapolation of the magnetic field in the region of interest. 
Panel a: side view showing the topology of the null point in  red-blue field lines colour-coded by field strength from 0.01–1~G; The light/dark green field lines are traced from the core of the radio isolines at 151~MHz and 432~MHz, as in panel~c; the light blue field lines are traced through the CME bubble, as in panel~e.
Panel~b: Earth-view showing the same null point blue-red field lines as in panel~a; contours in panel~b indicate the half-maximum NRH radio emission at 228~MHz (violet), 432~MHz (orange), and 151~MHz (cyan) at 09:47:49~UT assuming a height of 1.2~R$_\odot$ from disc centre (as discussed in the text).
Panel~c: Selected field lines through the radio contours, traced from the centre of orange radio contour (432~MHz) to the east (yellow field lines), the cyan isocontour (151) in the middle (light green field lines), and the orange isocontour (432~MHz) to the west.
Panel d: STEREO-A 171~\AA\ image at 09:46~UT \citep[processed using a Normalised Radial Gradient Filter, NRGF;][]{Morgan:2006} shown in blue/red colour table in the plane of sky. The red dots indicate the edge of the CME bubble identified in Figure~\ref{fig:wave_evol}.
Panel~e: PFSS field lines passing through the red dots from panel~d from a south-west prospective (same as in panel~a).  
\corrtwo{The magnetogram in panels~a-c shows the radial component of the magnetic field, with units of Gauss as indicated by the colourbar in panel~a.} In panels~d and e the same AIA 171~\AA\ subdomain as outlined by the white square in Figure~\ref{fig:radio_context}b is overlaid for reference.}\label{fig:pfss}}
\end{center}
\end{figure*}

These measurements indicate that although the plasma diagnostics of the different regions of interest provide a useful insight into the ability of the studied global shock to produce the observed radio emission, they do not tell the whole story. In particular, the global shock studied here appears to be much too weak in the identified locations to produce the observed type~\Rmnum{3} emission. However, the magnetic field through which the global shock propagated may provide some clues as to the origin of this emission.

\subsection{Magnetic field modelling}\label{ss:bfield}

To investigate the coronal magnetic field at the time of the eruption, a PFSS model was constructed using the \corr{Python} pfsspy package \citep{pfsspy_yeates,Stansby:2020}, with the JSOC daily synoptic map of 30 September 2011\footnote{\corr{hmi.mrdailysynframe\_small\_720s.20110930\_120000\_TAI.data.fits}} providing the photospheric boundary condition and the source surface placed at 4 solar radii ($R_s$). 
\corr{The pfsspy code was configured to internally remap the magnetogram using 360 nodes in longitude and 102 nodes in cosine-latitude, all uniformly spaced. In the radial direction, pfsspy employs a logarithmic grid, in this application using 120 nodes.}

Figure~\ref{fig:pfss} shows some of the field lines of the PFSS extrapolation, chosen using the field of view identified by the white box in Figure~\ref{fig:context}b. As shown from the side in Figure~\ref{fig:pfss}a\corr{, and as viewed from above in panel~b}, the region of interest is entirely surmounted by the fan of a high-altitude null point (at approximately 0.28 $R_s$), whose inner spine is rooted in the negative sunspot of the \corr{source} active region while the outer spine connects back to the photosphere (at the North East side). Such a strong topological constraint prevents any field line starting from within the area under the fan from connecting to interplanetary space. 

This PFSS extrapolation provides some insight into the origin of the observed radio emission. As the origin of the radio emission is a function of coronal density, it is not possible to know the exact location of the radio emission in the 3D extrapolation without making some assumptions. Firstly, we consider all the radio signals as coming from a single plane perpendicular to the line-of-sight at a height of \corr{around} 1.2~$R_s$ \corr{above the photosphere}. Next, the helioprojective coordinates of the radio maps were used to align them to the PFSS extrapolation as seen from Earth. It should be noted that this is a largely arbitrary approach as it is physically incorrect to assume that the radio signals are coming from a planar surface (the height of which was estimated using average coronal plasma properties) and that such a planar surface is the same for all frequencies. However, these assumptions enable us to make a qualitative validation of the emission scenario that includes both magnetic field and radio emission. Figure~\ref{fig:pfss}c shows the result of such a comparison with selected field lines under the fan (yellow) and close to the null but external to the fan (dark green). The isocontours show the half-maximum NRH radio emission at 432~MHz (orange), 228~MHz (violet), and 151~MHz (cyan) at 09:47:49~UT.

The emission corresponding to source~1 is associated with the field lines under the fan (yellow in Figure~\ref{fig:pfss}c) and is not connected to the magnetic null point, so the energetic particles are unable to escape into the heliosphere and instead are accelerated along closed magnetic loops. This matches the radio emission indicated by the first vertical dashed line in Figure~\ref{fig:radio_context}d, which was observed by NRH but not NDA (\cf\ the lack of a corresponding radio signature in Figure~\ref{fig:radio_context}e). In contrast, the type~\Rmnum{3} emission corresponding to source~2 and the active region source is connected to the magnetic null and hence open field lines (see dark green field lines in Figure~\ref{fig:pfss}a, c). As the CME expands against the spine, it activates reconnection at the null. As a result, the energetic particles accelerated by the shock at source~2 and by ongoing reconnection in the active region can escape along these open magnetic field lines into the heliosphere. This matches the radio emission associated with the second vertical dashed line in Figure~\ref{fig:radio_context}e which was observed by both NRH and NDA. The comparison between the CME bubble observed by STEREO and the field lines from the PFSS extrapolation (Figure~\ref{fig:pfss}d, e) show that the CME was still quite low when the type~\Rmnum{3} emission was observed; consistent with the observations of the EUV wave. These side-on observations enable us to identify which field lines were impacted by the CME contemporaneously with the production of the type~\Rmnum{3} radio signal. Figure~\ref{fig:pfss}d shows the size of the expanding bubble at 09:46~UT, the closest image available in time to the radio burst, identified by a series of \corr{red} dots, as viewed from the STEREO perspective. The corresponding PFSS field lines originating from those dots are shown in Figure~\ref{fig:pfss}e as seen from a S-E perspective, \corr{and in Figure~\ref{fig:pfss}a too for comparison with the fan structure}. 

Although the PFSS extrapolation does a good job at describing the static structure of the surrounding coronal magnetic field, it does not account for reconfiguration of the coronal magnetic field during an eruption. However, the combination of radio emission and PFSS extrapolation does provide a unique insight into the configuration of the coronal magnetic field at the time of this eruption. In particular, the spine of the fan topology observed above the active region here enables particles accelerated by reconnection in the active region to escape into the heliosphere, resulting in the observed extended type~\Rmnum{3} emission. The first type~\Rmnum{3} burst (corresponding to the 1st burst observed ahead of the EUV wave) is produced by particles accelerated by the shock along closed magnetic field lines, while the second burst observed ahead of the EUV wave corresponds to the footpoints of magnetic loops connected to the null. 

\section{Discussion and conclusions}\label{s:disc}

As a well-observed, albeit weak, solar eruptive event with a CME, global shock wave, and associated radio emission, this event provides an excellent opportunity to study the acceleration of energetic particles by a weak shock. The global shock wave was tracked and analysed using an intensity profile approach, and propagated anisotropically with a velocity of $\sim$400-500~km~s$^{-1}$ and a corresponding Alfv\'{e}n Mach number of 1.001--1.013. In spite of this, it was able to accelerate electrons low down in the corona which produced distinct type~\Rmnum{3} emission. However, the radio emission was highly localised, suggesting that only certain locations along the shock front achieved the required criteria for accelerating the energetic particles. 

Global waves in the low solar corona have traditionally been interpreted as propagating through a radial coronal magnetic field, and are often therefore interpreted as quasi-perpendicular shocks \citep[\cf][]{Carley:2013}. This quasiperpendicularity enables efficient acceleration of electrons via shock drift acceleration \citep[SDA;][]{Ball:2001}, which involves adiabatic reflection of particles from the shock, with each reflection increasing the energy of the particle. For coronal shocks, the multiple reflection process has been explained using inhomogeneities in the shock front (``ripples''), with this quasiperiodic structure thought to explain the presence of herringbones in type~\Rmnum{2} and \Rmnum{3} radio emission \citep[\eg][]{Carley:2013,Carley:2015,Morosan:2019}. Here, source~1 appears to have a herringbone structure, but source~2 does not, suggesting a very limited deformation of the shock front assuming this interpretation.

The radio emission at both sources identified in Figure~\ref{fig:radio_context}a, b can be clearly identified to be ahead of the leading edge of the global EUV shock wave. \corr{This suggests} that the radio emission observed here is due to a shock wave propagating ahead of the observed EUV front\corr{. In this case, the EUV observations provide a lower limit on the estimated kinematics of the shock front, as the shock must heat the plasma sufficiently to produce the observed perturbation \citep[\cf][]{Ma:2011}. We can then estimate the shock speed at $\sim1.35$ times the observed EUV wave speed, assuming the shock starts at the same time and location as the EUV wavefront and reaches the centroid of the observed radio source at the time of the observed radio emission.}

To \corr{study this in more detail}, both the plasma environment in the region through which the global shock wave evolved and the coronal magnetic field were examined using a regularised inversion DEM approach and a PFSS extrapolation respectively. First, the plasma evolution was examined in the two regions identified as being the sources locations for the type~\Rmnum{3} emission and a third region chosen as a background reference. All three regions were shown to exhibit increases in temperature and density associated with the passage of the global shock wave (see Figure~\ref{fig:temp_dens_evol}). The variation in density and temperature measured in each region is proportional to the distance from the source active region, with the distinct, strong increases in both temperature and density observed in region~3 most likely due to blobs of plasma ejected by the active region passing through the field of view.

The DEM also enabled a detailed examination of the properties of the global EUV shock wave at the three regions of interest. In each location, the global EUV shock wave was found to be very weak, with a hydrodynamic Mach number ranging from 1.9 in region~1 to $\sim$1.4 in regions~2 \& 3. However, the Alfv\'{e}n Mach number was found to be much lower, ranging from $\sim$1.008 (1.01) for an angle of $\theta=0^{\circ}$ ($\theta=90^{\circ}$) between the magnetic field and shock normal in regions~1 \& 2 to $\sim$1.001 (1.003) for an angle of $\theta=0^{\circ}$ ($\theta=90^{\circ}$) between the magnetic field and shock normal in region~3. This indicates that while the global shock wave was stronger in the locations corresponding to the type~\Rmnum{3} radio emission, it was still very weak, and most likely could not have produced the particle acceleration observed here.

The surrounding coronal magnetic field was then examined using a PFSS extrapolation. This investigation revealed a coronal null point above the \corr{source} active region as shown in Figure~\ref{fig:pfss}a, which suggests that the eruption occurred in a closed magnetic topology without direct access to magnetic field opening into the heliosphere. However, the kinematic evolution of the erupting CME was found to be best fit by a quadratic function, \corr{kinematically} consistent with an eruption evolving via a breakout model scenario \citep[\cf][]{Antiochos:1999}. The first type~\Rmnum{3} emission corresponding to source~1 (\cf\ Figure~\ref{fig:radio_context}a) was found to be due to particles accelerated both upwards and downwards along a closed magnetic loop, as evidenced by the bifurcated emission (Figure~\ref{fig:radio_context}f) and lack of low frequency emission associated with the burst. In contrast, the radio emission associated with source~2 (\cf\ Figure~\ref{fig:radio_context}b) was due to particles accelerated upwards away from the Sun along open magnetic field lines, as evidenced by the lack of a bifurcated signature in the NRH spectrum (Figure~\ref{fig:radio_context}g) and the low frequency emission observed by the NDA.

\corr{The event can be initially described as follows. Starting at 09:47~UT, we note a temporal correlation between the eruption and the observed radio signal, which includes type~\Rmnum{3} bursts, indicating access to open magnetic field. However, the source active region (AR~11305) is isolated from the open field, as it is fully surmounted by the fan of a (relatively) high-altitude null point, the inner spine of which is rooted in the active region. The earliest observation of the CME bubble by \emph{STEREO} shows the bubble expanding but still far from the null point, with the only open field lines from the PFSS extrapolation rooted at the edge of the fan of the null, in the positive polarity of the neighbouring AR~11306. Due to the proximity to the fan, these field lines spread out continuously from the spine of the null to the open field. We can propose the following scenario to explain these observations. The CME bubble expands in the SE direction, but also against the spine of the null rooted in the negative polarity of AR~11305. This activates reconnection at the null, impacting the dark-green field lines shown in Figure~\ref{fig:pfss}a, and eventually providing access to open field. This scenario is feasible as there are multiple ways for reconnection at the null to provide access to the dark-green field lines in Figure~\ref{fig:pfss}a. These include null deformation due to the expanding CME bubble \citep[\cf][]{Pontin:2007,Galsgaard:2011}, or slipping reconnection across the outer spine quasi-separatrix layer at the null \citep[\eg][]{Masson:2017}. However, this hypothesis is limited by the potential nature of the PFSS extrapolation, which means that we cannot pinpoint a specific mechanism. As a result, we must limit our suggestion to a generic ``activation'' of the null point driven by the expanding CME bubble}

The combination of ongoing reconnection in the \corr{source} active region and the compression of the magnetic spine rooted in the active region by the erupting CME then resulted in continuous acceleration of energetic particles, producing the long lasting type~\Rmnum{3} emission identified as the AR source in Figure~\ref{fig:radio_context}c. 

\corr{Therefore}, rather than the bursty emission resulting from the repeated acceleration of electrons by the rippled front of the global shock wave as previously observed by \citet{Carley:2013,Morosan:2019}, the radio emission observed here was produced by electrons accelerated by the interaction of the global shock with the surrounding coronal magnetic field. Although this scenario might be expected for every CME erupting into the surrounding coronal magnetic field, the very weak nature of the eruption and shock observed here enabled the properties of the global EUV wave and the associated spatially-resolved type~\Rmnum{3} radio emission to be disentangled from the main eruption and studied in detail for the first time. The ability of \emph{Solar Orbiter} and \emph{Parker Solar Probe} to observe the solar corona at higher spatial and temporal resolution than currently possible and detect solar energetic particles closer to their source will enable a more thorough investigation and hence deeper understanding of this phenomenon.

\acknowledgments
\corr{We wish to thank the anonymous referee whose comments helped to improve the article.} DML is grateful to the Science Technology and Facilities Council for the award of an Ernest Rutherford Fellowship (ST/R003246/1), and thanks ISSI (International Space Science Institute, Bern) for the hospitality provided to the team ``Foreshocks Across the Heliosphere'' led by H.~Hietala and F.~Plaschke, whose discussions helped to clarify the ideas described in this paper. G.V. acknowledges the support  from the European Union's Horizon 2020 research and innovation programme under grant agreement No 824135 and of the STFC grant number ST/T000317/1. JO is supported by funding from the Science and Technology Facilities Council (STFC) studentship ST/R505171/1.
\vspace{5mm}\\
\facilities{SDO, NRH, STEREO, NDA}
\software{SolarSoftWare \citep{Freeland:1998}; pfsspy \citep{pfsspy_yeates,Stansby:2020}}

\bibliographystyle{aasjournal}
\bibliography{references}{}

\end{document}